\documentstyle[emulateapj,psfig]{article}
\begin{document}

\title{GeV Emission from TeV Blazars and Intergalactic Magnetic Fields}
\author{Z. G. Dai$^{1,2}$, B. Zhang$^2$, L. J. Gou$^2$, P. M\'esz\'aros$^{2,3}$ 
        and E. Waxman$^4$}
\affil{$^1$Department of Astronomy, Nanjing University, Nanjing
           210093, China\\
       $^2$Dept. Astronomy \& Astrophysics, 525 Davey Lab, Pennsylvania State
           University, University Park, PA 16802\\
       $^3$Dept. of Physics, 102 Davey Lab., Pennsylvania State
            University, University Park, PA 16802\\
       $^4$Department of Condensed Matter Physics, Weizmann Institute,
           Rehovot 76100, Israel}

\begin{abstract}
Several high-frequency peaked BL Lac objects such as Mrk 501 are
strong TeV emitters. However, a significant fraction of the TeV
gamma rays emitted are likely to be absorbed in
interactions with the diffuse IR background, yielding
electron-positron pairs. Hence, the observed TeV spectrum
must be steeper than the intrinsic one. Using the recently derived
intrinsic $\gamma$-ray spectrum of Mrk 501 during its 1997 high
state, we study the inverse-Compton  scattering of cosmic
microwave photons by the resulting electron-positron pairs, which
implies the existence of a hitherto undiscovered GeV emission. The
typical duration of the GeV emission is determined by
the flaring activity time and the energy-dependent magnetic
deflection time. We numerically calculate the scattered photon
spectrum for different intergalactic magnetic field (IGMF)
strengths, and find a spectral turnover and flare duration at GeV
energies which are dependent on the field strength. We also
estimate the scattered photon flux in the quiescent state of Mrk
501. The GeV flux levels predicted are consistent with existing
EGRET upper limits, and should be detectable above the synchrotron
-- self Compton (SSC) component with the {\em Gamma-Ray Large Area Space
Telescope} ({\em GLAST}) for IGMFs $\lesssim 10^{-16}$ G, as
expected in voids. Such detections would provide constraints on
the strength of weak IGMFs.
\end{abstract}

\keywords{BL Lacertae objects: general --- BL Lacertae objects:
individual (Markarian 501) --- diffuse radiation
--- gamma rays: theory --- magnetic fields }

\section{Introduction}

Blazars including high-frequency peaked BL Lac objects (HBLs) are
the most extreme and powerful sources among active galactic
nuclei. The standard blazar model consists of a supermassive black
hole ejecting twin relativistic jets, one of which is close to the line of sight. 
Several HBLs such as Mrk 501, Mrk 421, PKS 2155-304, 1ES 2344$+$514, 
H1426$+$428 and 1ES1959$+$650 are of particular interest because 
they emit TeV photons (see Catanese \& Weekes 1999; Horns et al. 2002). 
The detection and study of such photons can provide new insights on
the energetics and physical conditions in the emission regions of
such blazars (Katarzy\'nski, Sol \& Kus 2001; Kino, Takahara \& 
Kusunose 2002). Also, constraints on the spectral energy distribution
of the intergalactic infrared background can be inferred from the
 observations on TeV photons (for a review see Hauser \& Dwek 2001).
Stecker, De Jager \& Salamon (1992) have emphasized that the
high-energy gamma photon spectra from these blazars will be modified
by strongly redshift-dependent absorption effects due to interactions
of such photons with the intergalactic infrared-UV background, and
indicated that the  intrinsic spectrum of  an observed TeV blazar
can be derived by evaluating the optical depth to TeV photons. Such
calculations were made for Mrk 501 during the 1997 flaring activity,
leading to an inferred intrinsic high-energy spectrum with a broad,
flat peak that is much higher than the observed one in the $\sim 5-10$
TeV range (Konopelko et al. 1999; De Jager \& Stecker 2002, hereafter
DS). The physical reason for this difference is that a significant
fraction of the original high-energy gamma rays have been absorbed
in $\gamma\gamma$ interactions with photons of the intergalactic
infrared-UV background, leading to electron/positron pairs.

The purpose of this Letter is to suggest that inverse Compton (IC)
scattering of the resulting electron/positron pairs against cosmic
microwave background (CMB) photons may produce a new GeV emission
component in TeV blazars. For gamma-ray bursts, such Compton scattering
leads to an observable, delayed MeV-GeV emission component if the
intergalactic magnetic fields (IGMFs) are very weak (Plaga 1995;
Cheng \& Cheng 1996; Dai \& Lu 2002). A similar phenomenon is also
expected from gamma-ray burst proton interactions with the CMB
(Waxman \& Coppi 1996). Here we discuss the well-studied blazar Mrk
501, both because the high-energy spectrum up to 20 TeV of strong
flares of this blazar in 1997 has been observed by the HEGRA air
Cerenkov telescope system (Aharonian et al. 1999, 2001), and
because the intrinsic spectrum of Mrk 501 over two decades of
energy has been derived based on the consistency between the
Whipple telescope and HEGRA spectra (DS).

The strength of IGMFs has not been determined so far. Faraday
rotation  measures imply an upper limit of $\sim 10^{-9}$ G for a
field with  1 Mpc correlation length (Kronberg 1994 for a review).
Other methods were proposed to probe fields in the range
$10^{-10}$  G to $10^{-20}$ G (Lee, Olinto \& Sigl 1995; Plaga
1995). To interpret the observed $\mu$G magnetic fields in
galaxies and X-ray clusters, the seed fields required in dynamo
theories could be as low as $10^{-20}$ G (Kulsrud et al. 1997;
Kulsrud 1999). Furlanetto \& Loeb (2001) argued that quasar
outflows may pollute the intergalactic medium, but the possible
weak IGMFs in voids may remain  uncontaminated. Theoretical
calculations of primordial magnetic fields show that these fields
could be of order $10^{-20}$ G or even as low as $10^{-29}$ G,
generated during the cosmological QCD or electroweak phase
transition respectively (Sigl, Olinto \& Jedamzik 1997). In this
Letter we propose that by observing a hitherto undiscovered GeV
emission component from flares of TeV blazars such as Mrk 501, one
may be able to obtain important information or constraints on the
poorly known IGMFs.

\section{Properties of External IC Emission}

We consider a strong flare, e.g. in Mrk 501, of duration $t_{\rm
var}$. A fraction of the high-energy $\gamma$-rays emitted during
such a flare can be absorbed in the cosmic background radiation
fields as these $\gamma$-rays travel towards the observer. The
pair production optical depth $\tau_{\gamma\gamma}^{\rm ex}$
depends strongly on the $\gamma$-ray energy ($E_\gamma$) and the
redshift ($z$). DS numerically calculated
$\tau_{\gamma\gamma}^{\rm ex}$ as a function of the photon energy
for low redshifts by using the models of Malkan \& Stecker (2001)
of the infrared background radiation, and extrapolating these
models into the optical-UV range in terms of recent galaxy count
data. DS considered both ``fast evolution" and ``baseline" cases,
which may be considered to bracket the spectral energy
distribution of the intergalactic infrared-UV background
radiation. If primary photons of energy $E_\gamma$ are absorbed,
the resulting electron/positron pairs have Lorentz factors
$\gamma_e \simeq E_\gamma/(2m_ec^2)= 10^6(E_\gamma/1\,{\rm TeV})$,
where $m_e$ is the electron mass. The pairs will subsequently
Compton scatter on the ambient CMB photons. As a result, the
initial energy of a microwave photon, ${\bar\epsilon}$, is boosted
by IC scattering up to an average value $\sim
\gamma_e^2{\bar\epsilon}\simeq 0.63(E_\gamma/1\,{\rm TeV})^2$ GeV,
where ${\bar\epsilon}=2.7kT$ is the mean energy of the CMB photons
with $T\simeq 2.73\,$K and $k$ is the Boltzmann constant.

\subsection{The GeV Emission Duration}

Several timescales are involved in the emission process. The first
is the well-known angular spreading time,
$\Delta t_{\rm A}\simeq R_{\rm pair}/(2\gamma_e^2c)=
96(\gamma_e/10^6)^{-2}(n_{\rm IR}/1\,{\rm cm}^{-3})^{-1}\,{\rm s}$,
where $R_{\rm pair}=(0.26\sigma_Tn_{\rm IR})^{-1}\simeq 5.8\times
10^{24}(n_{\rm IR}/1\,{\rm cm}^{-3})^{-1}$ cm is the typical
pair-production distance, $\sigma_T$ is the Thomson cross section,
and $n_{\rm IR}$ is the intergalactic infrared photon number
density (see Dai \& Lu 2002). Therefore, primary TeV photons have 
a typical mean free path of a few Mpc, so that although the source may 
be in a region of a high field with $\sim 10^{-9}$ G, these photons may 
escape to much lower field regions.  Six TeV blazars detected so far 
have redshifts of $\sim 0.03$ to $\sim 0.1$, implying that their luminosity 
distances are much larger than $R_{\rm pair}$. Only in such cases are 
our calculations valid. 

The IC cooling timescale (in the local rest frame) of relativistic
electrons with Lorentz factor of $\gamma_e$ is 
$t_{\rm IC} = 3m_ec/(4\gamma_e\sigma_Tu_{\rm cmb})= 7.3
\times 10^{13}(\gamma_e/10^6)^{-1}\,{\rm s}$,
where $u_{\rm cmb}$ is the CMB energy density. Thus the typical
flight path, in which most of the electron energy is lost, is
$\lambda_{\rm IC}\simeq ct_{\rm IC} =2.2\times
10^{24}(\gamma_e/10^6)^{-1}$ cm. This is much less than the
distance from the source to the observer, implying that energy
loss of the electron due to IC scattering is local. In the absence
of any IGMF, the IC cooling timescale in the observer frame would
be $\Delta t_{\rm IC}\simeq t_{\rm IC}/(2\gamma_e^2)
=37(\gamma_e/10^6)^{-3}\,{\rm s}$.

In the presence of IGMFs with strength $B_{\rm IG}$, the electrons
will be deflected. The deflection angle is estimated by
$\theta_{\rm B}\simeq \lambda_{\rm IC}/R_{\rm L}=1.3\times
10^{-5}(\gamma_e/10^6)^{-2}(B_{\rm IG}/10^{-20}{\rm G})$, where
$R_{\rm L}=\gamma_em_ec^2/(eB_{\rm IG})$ is the Larmor radius of
the electrons. The contribution to the emission time due to
magnetic deflection becomes $\Delta t_{\rm B} \simeq (1/2)t_{\rm IC}
\theta_{\rm B}^2\simeq 6.1\times 10^3(\gamma_e/10^6)^{-5}(B_{\rm
IG}/10^{-20}{\rm G})^2\,{\rm s}$ for $\theta_{\rm B}\ll 1$. 
This may provide a probe of weak IGMFs
and its application to the delayed high-energy emission of GRB
940217 yields IGMFs which are as weak as $B_{\rm IG}\sim
10^{-20}$\,G.

Therefore, the duration estimate of the IC emission from
electron/positron pairs scattering off the CMB is 
$\Delta t (\gamma_e) = \max(\Delta t_{\rm IC}, \Delta t_{\rm A},
\Delta t_{\rm B}, t_{\rm var})$. 
Taking $\Delta t_{\rm B}= t_{\rm var}$, 
which implies a scattered photon energy 
$\sim E_{\rm turn}= 0.22(B_{\rm IG}/10^{-20}{\rm G})^{4/5}
(t_{\rm var}/1\,{\rm day})^{-2/5}\,{\rm GeV}$, we find that 
the duration of the expected GeV emission is always given by 
the variability timescale of the TeV gamma-ray flux for photon 
energes larger than $E_{\rm turn}$, and it is given by 
$\Delta t_{\rm B}$ for photon energies smaller than $E_{\rm turn}$.   

\subsection{The Emission Spectrum}

The optical depth ($\tau_{\gamma\gamma}^{\rm ex}$) to high-energy
photons was used by DS to derive the intrinsic $\gamma$-ray spectrum
of Mrk 501 during its 1997 high state, which can be parameterized as
\begin{equation}
E_\gamma^2\frac{dN_\gamma}{dE_\gamma}=KE_\gamma^{2-\Gamma_1}
\left[1+\left(\frac{E_\gamma}{E_B}\right)^f\right]^{(\Gamma_1-\Gamma_2)/f},
\end{equation}
where $E_\gamma=1(\gamma_e/10^6)$ TeV, and the parameters $E_B$
(in TeV), $K$ (in $10^{-10}\,{\rm ergs}\,{\rm cm}^{-2}\,{\rm
s}^{-1}$), $\Gamma_1$, $\Gamma_2$ and $f$ are shown in Table 3 of DS. 
Please note that equation (1) is a time-averaged spectrum of a long-term 
outburst of about 6 months in 1997. During this period Mrk 501 has shown 
a number of one-day flares, whose gamma-ray flux values appear to be larger than 
the time-averaged ones by a factor of $\eta$ with $1 \lesssim \eta \lesssim 3$. 
Thus, for accurate calculations of the GeV emission one should use 
the specific flux values for the flare considered. Here we take $\eta =1$ 
as a conservative value.   Letting the luminosity distance to 
the source be $D_L\simeq 5\times 10^{26}$ cm, we have the total electron 
energy spectrum (including the positrons),
\begin{equation}
\frac{dN_e}{d\gamma_e} =
CE_\gamma^{-\Gamma_1}\left[1+\left(\frac{E_\gamma}{E_B}
\right)^f\right]^{(\Gamma_1-\Gamma_2)/f}
\left(1-e^{-\tau_{\gamma\gamma}^{\rm ex}}\right),
\end{equation}
where $C=5\pi\times 10^{-6}D_L^2K$. Therefore, the observed
time-averaged scattered-photon spectrum is given by (Blumenthal \&
Gould 1970)
\begin{equation}
\frac{dN_\gamma^{\rm SC}}{dE_{\gamma,1}}=\frac{1}{4\pi
D_L^2}\int\int \left(\frac{dN_e}{d\gamma_e}\right)
\left(\frac{dN_{\gamma_e,\epsilon}}{dtdE_{\gamma,1}}\right)t_{\rm
IC}\xi d\gamma_e,
\end{equation}
where $\xi\equiv t_{\rm var}/\Delta t$, $E_{\gamma,1}$ is the
externally scattered photon energy, $t$ is the time measured in
the local rest frame, and
$dN_{\gamma_e,\epsilon}/dtdE_{\gamma,1}$, expressed by equation
(2.48) of Blumenthal \& Gould (1970) using the Klein-Nishina (KN)
cross-section formula, is the spectrum of photons scattered by an
electron with Lorentz factor of $\gamma_e$ from a segment of the
CMB photon gas of differential number density
$n(\epsilon)d\epsilon$. If the electron energy spectrum is
simplified as $\propto \gamma_e^{-\Gamma_1}$, then a primary
analysis similar to Dai \& Lu (2002) shows that in the Thomson
limit the scattered photon spectrum $\propto
E_{\gamma,1}^{-(\Gamma_1+2)/2}$ for $\Delta t_{\rm B}(\gamma_e)\ll
t_{\rm var}$ and $\propto E_{\gamma,1}^{-(\Gamma_1-3)/2}$ for
$\Delta t_{\rm B}(\gamma_e)\gg t_{\rm var}$, implying a spectral
turnover at scattered photon energies of $\sim E_{\rm turn}$. 
Figure 1 shows the energy spectra of scattered photons 
for different assumed IGMFs during the 1997 flaring activity of Mrk 501. 
For an IGMF $\gtrsim 10^{-20}$ G, we indeed see a spectral turnover,
whose energy is strongly dependent on the field strength. Figure 2 
presents the emission spectra for different flare durations, and we find
that for a fixed IGMF, the longer the duration, the smaller the turnover 
energy becomes and thus the emission becomes easier to detect.

\section{Comparison with the SSC Spectrum}

The observed low-energy hump of Mrk 501 is usually thought to be
due to synchrotron radiation from a relativistic plasma blob. The
comoving spectral synchrotron photon energy density is given by
$u'(\epsilon'_\gamma)=[2 D_L^2\phi(\epsilon_\gamma)]
/(r_b^2c\delta^4)$, where $\phi(\epsilon_\gamma)$ is the observed
synchrotron flux at energy of $\epsilon_\gamma$, $r_b$ is the
comoving radius, $\delta$ is the Doppler factor, and primes denote
comoving quantities. Here $\epsilon_\gamma=\delta\epsilon'_\gamma$,
and $r_b=\delta ct_{\rm var}$.

In order to calculate the comoving spectral synchrotron photon
number density $n'(\epsilon'_\gamma)$, we approximate the flux
density $F(\epsilon_\gamma)\propto \epsilon_\gamma^{-\alpha_1}$
for $\epsilon_\gamma\le\epsilon_{\rm br}$ and
$F(\epsilon_\gamma)\propto \epsilon_\gamma^{-\alpha_2}$ for
$\epsilon_{\rm br}<\epsilon_\gamma\le \epsilon_{\rm pk}$ observed
during the flares of 1997. From the observations by Catanese et
al. (1997), Pian et al. (1998) and Petry et al. (2000), it can be
seen that $\alpha_1\simeq 0.4$, $\alpha_2\simeq 0.6$,
$\epsilon_{\rm br}\simeq 2$ keV and $\epsilon_{\rm pk}\sim 100$
keV, and the synchrotron radiation flux
$\phi(\epsilon_\gamma)\equiv \epsilon_\gamma F(\epsilon_\gamma)$
reaches a maximum value $\phi_{\rm pk}\sim 6\times 10^{-10}\,{\rm
ergs}\,{\rm cm}^{-2}\,{\rm s}^{-1}$ at $\epsilon_\gamma
=\epsilon_{\rm pk}$.

\subsection{The Intrinsic Optical Depth}

We first derive a generic formula for the optical depth to a
photon with $E_\gamma>1$\,TeV inside the blob. In the comoving
frame of the blob, this photon can annihilate a second softer
photon with energy greater than $\epsilon'_{\rm
an}=\delta(m_ec^2)^2/E_\gamma$, yielding an electron/positron
pair. The comoving total number density of soft photons that have
energies greater than $\epsilon'_{\rm an}$ is calculated by
\begin{equation}
n'(>\epsilon'_{\rm an})=\frac{2D_L^2\phi_{\rm
pk}}{\alpha_1r_b^2c\delta^4}\left(\frac{\epsilon_{\rm
br}}{\epsilon_{\rm pk}}\right)^{-\alpha_2+1}
\left(\frac{\delta}{\epsilon_{\rm
br}}\right)^{-\alpha_1+1}\epsilon_{\rm an}^{\prime -\alpha_1},
\end{equation}
for $\epsilon'_{\rm an}<\epsilon_{\rm br}/\delta$. Thus, the
optical depth is $\tau_{\gamma\gamma}^{\rm in}=(11/180)
\sigma_Tn'(>\epsilon'_{\rm an})r_b$, where $(11/180)\sigma_T$ 
is the averaged cross section assuming an $F(\epsilon_\gamma)
\propto \epsilon_\gamma^{-1}$ spectrum (Svensson 1987; 
also see Lithwick \& Sari 2001 and Zhang \& M\'esz\'aros 2001). 
Inserting equation (4) into the above
equation yields
\begin{equation}
\tau_{\gamma\gamma}^{\rm in}={\hat
\tau}(E_\gamma/m_ec^2)^{\alpha_1}\delta^{-2\alpha_1-4},
\end{equation}
where
\begin{equation}
{\hat \tau}=\frac{(11/90)\sigma_TD_L^2(\phi_{\rm pk}/\epsilon_{\rm
pk})(\epsilon_{\rm pk}/\epsilon_{\rm br})^{\alpha_2}(\epsilon_{\rm
br}/m_ec^2)^{\alpha_1}}{\alpha_1c^2t_{\rm var}}.
\end{equation}
For $\alpha_1=\alpha_2=1$, equation (5) becomes
$\tau_{\gamma\gamma}^{\rm in}\propto \delta^{-6}$, which is
consistent with the previously used optical depth (see Aharonian
et al. 1999; DS). Inserting the observed values of the parameters
into equations (5) and (6), we have $\tau_{\gamma\gamma}^{\rm
in}=0.015(E_\gamma/1\,{\rm TeV})^{0.4}(\delta/10)^{-4.8}(t_{\rm
var}/1\,{\rm day})^{-1}$. The observed highest energy
$E_\gamma\sim 20$ TeV implies that the optical depth to a photon
with such an energy is less than unity. This provides a lower
limit to the Doppler factor $\delta\ge 5.4(E_\gamma/20\,{\rm
TeV})^{1/12}(t_{\rm var}/1\,{\rm day})^{-1/4.8}$, i.e., the
photon-energy cutoff due to pair production in the blob can be
larger than 20 TeV for $\delta>5.4$.

\subsection{The SSC Spectrum}

We next discuss the SSC spectrum from the blob. We assume the
accelerated electrons in the blob to have a power-law energy
distribution with an index of $p$ and the maximum Lorentz factor
of $\gamma_{\rm M}$. We also assume that the magnetic field
strength in the blob is $B'$. Since the synchrotron peak of the
$\phi(\epsilon_\gamma)$ flux corresponds to the emission from
electrons with the maximum Lorentz factor, we obtain $\gamma_{\rm
M}=2.93\times 10^6(\delta/10)^{-1/2}(B'/0.1\,{\rm
G})^{-1/2}(\epsilon_{\rm pk}/100\,{\rm keV})^{1/2}$. Because of
$\gamma_{\rm M}\epsilon_{\rm br}\gg \delta m_ec^2$, relativistic
KN corrections to the Compton spectrum are very important. We
estimate the lower limit to the observed gamma energy at which the
KN effects should be considered, $E_{\rm KN}\sim (\delta
m_ec^2)^2/\epsilon_{\rm br}=13(\delta/10)^2(\epsilon_{\rm
br}/2\,{\rm keV})^{-1}$ GeV. In addition, the Compton peak energy
in the KN range is approximated by $E_{\gamma,{\rm pk}} \sim
\delta \gamma_{\rm M}m_ec^2= 1.5(\delta/10)^{1/2}(B'/0.1\,{\rm
G})^{-1/2}(\epsilon_{\rm pk}/100\,{\rm keV})^{1/2}\,{\rm TeV}$.
Thus, the SSC photon spectrum  
 $ \propto E_\gamma^{-(p+1)/2}$ for $E_\gamma\le E_{\rm KN}$ and 
$\propto E_\gamma^{-p}$ for $E_{\rm KN}<E_\gamma\le 
E_{\gamma,{\rm pk}}$. Comparing this spectrum with equation (1), 
we find $p=\Gamma_1\simeq 1.6-1.7$, which is not only consistent with 
the value of $p$ shown by Kino et al. (2002) but is also in agreement
with the index of $\alpha_2$ obtained from the BeppoSAX
observations by Pian et al. (1998).

Figure 1 also presents the SSC spectra, whose high-energy segments
are assumed to be the intrinsic spectra derived by DS. It can be
seen from this figure that the flux of the externally scattered
photons is higher than that of the SSC photons in the blob for an
IGMF $\lesssim 10^{-16}$ G. Furthermore, the externally scattered
photon flux is consistent with the existing observations by EGRET
and large enough to be detected by {\em GLAST}.

\section{Discussion and Conclusions}

Using the intrinsic $\gamma$-ray spectrum of Mrk 501 during its
1997 flaring activity derived by DS as an example, we have
predicted a new GeV emission component for this and similar HBLs,
if their radiation reaches us through regions with a low magnetic
field as expected in under-dense voids. This GeV emission is due
to the IC scattering of CMB photons by the electron/positron pairs
produced in interactions of high-energy photons with the cosmic
infrared-UV background photons. We summarize our findings: First,
the GeV flux of the scattered CMB photons is higher
than that from the SSC process in the blob for IGMFs $\lesssim
10^{-16}$\,G. For an IGMF $\gtrsim 10^{-20}$ G, there is an
observable spectral turnover, whose position is strongly dependent
on the field strength. This spectral turnover is due to the fact
that the lower energy emission would have a longer delay. Second,
the typical duration of the GeV emission is always given by 
the variability timescale of the TeV gamma-ray flux for photon 
energes larger than $E_{\rm turn}$, and it is given by 
$\Delta t_{\rm B}\propto E_{\gamma,1}^{-5/2}$ for photon 
energies smaller than $E_{\rm turn}$. Third, the GeV emission
predicted here is consistent with existing EGRET observations, and
is strong enough to be detected by {\em GLAST}, for IGMFs
$\lesssim 10^{-16}$ G and flare times $\sim 1$ day.
Fourth, we have derived a generic formula for the optical depth
due to pair production in a relativistic blob,
$\tau_{\gamma\gamma}^{\rm in}\propto \delta^{-2\alpha-4}$, where
$\delta$ is the Doppler factor and $\alpha$ is the index of the
softer photon energy spectrum. This optical depth is not only 
consistent with the previously used formula in the $\alpha=1$ case  
but can also be generalized to broader cases.  

We note that even if a source such as Mrk 501 is not in a void, primary 
TeV photons emitted from the source have such a long mean free path
that electron/positron pairs may be produced in void regions.
Thus, detections on the GeV emission would provide a sensitive
probe for weak IGMFs, of consequence for cosmogonical and galactic
dynamo theories.

Our results are also relevant for the quiescent state of HBLs such
as Mrk 501. Aharonian, Coppi, \& V\"olk (1994) and Coppi \&
Aharonian (1997) discussed a similar model for very high energy
emission ($>100$ GeV) from blazars in the quiescent state. We here
use the recent models for the intergalactic infrared-UV background
radiation to discuss lower energy emission from TeV blazars.
According to Catanese et al. (1997), the Whipple telescope
detected a flux of $\phi(E_\gamma\simeq 1\,{\rm TeV})=
8^{+2}_{-3}\times 10^{-12}\,{\rm ergs}\,{\rm cm}^{-2}\,{\rm
s}^{-1}$ in the quiescent state of this HBL. The external
pair-production optical depth $\tau_{\gamma\gamma}^{\rm
ex}(E_\gamma=1\,{\rm TeV})\simeq 0.5$ and $0.6$, and thus the flux
of the externally scattered photons with an energy of $\sim
0.6$ GeV is estimated as $\phi(E_\gamma\simeq 0.6\,{\rm
GeV})\simeq \phi(E_\gamma\simeq 1\,{\rm TeV})\times [\exp
(\tau_{\gamma\gamma}^{\rm ex})-1]\sim 5 \times 10^{-12}\,{\rm
ergs}\,{\rm cm}^{-2}\,{\rm s}^{-1}$ and $7\times 10^{-12}\,{\rm
ergs}\,{\rm cm}^{-2}\,{\rm s}^{-1}$ for the ``baseline" and
``fast-evolution" models of DS, respectively. These values of the
GeV emission flux are much larger than $\sim 1.0\times
10^{-13}\,{\rm ergs}\,{\rm cm}^{-2}\,{\rm s}^{-1}$, the
sensitivity of {\em GLAST} at energy $\sim 0.6$ GeV for
steady sources in a one-year survey. Therefore, even in the quiescent
state of Mrk 501, its externally scattered emission may also be
detected by {\em GLAST}. It should be pointed out that such detections
are possible for a magnetic deflection angle that is less than the
blazar jet opening angle. This also requires that the field
strength be below $\sim 10^{-16}$ G, similar to the flare case,
for an opening angle $\sim 0.1$. Above this value of the magnetic
field, the steady GeV emission may be suppressed.

\acknowledgments

We would like to thank the referee for valuable comments. This work 
was supported by NSF AST 0098416, NASA NAG5-9192, NAG5-9153,
and by the National Natural Science Foundation of China (grant 19825109)
and the National 973 Project (NKBRSF G19990754).

\begin{figure}
\centerline{\psfig{file=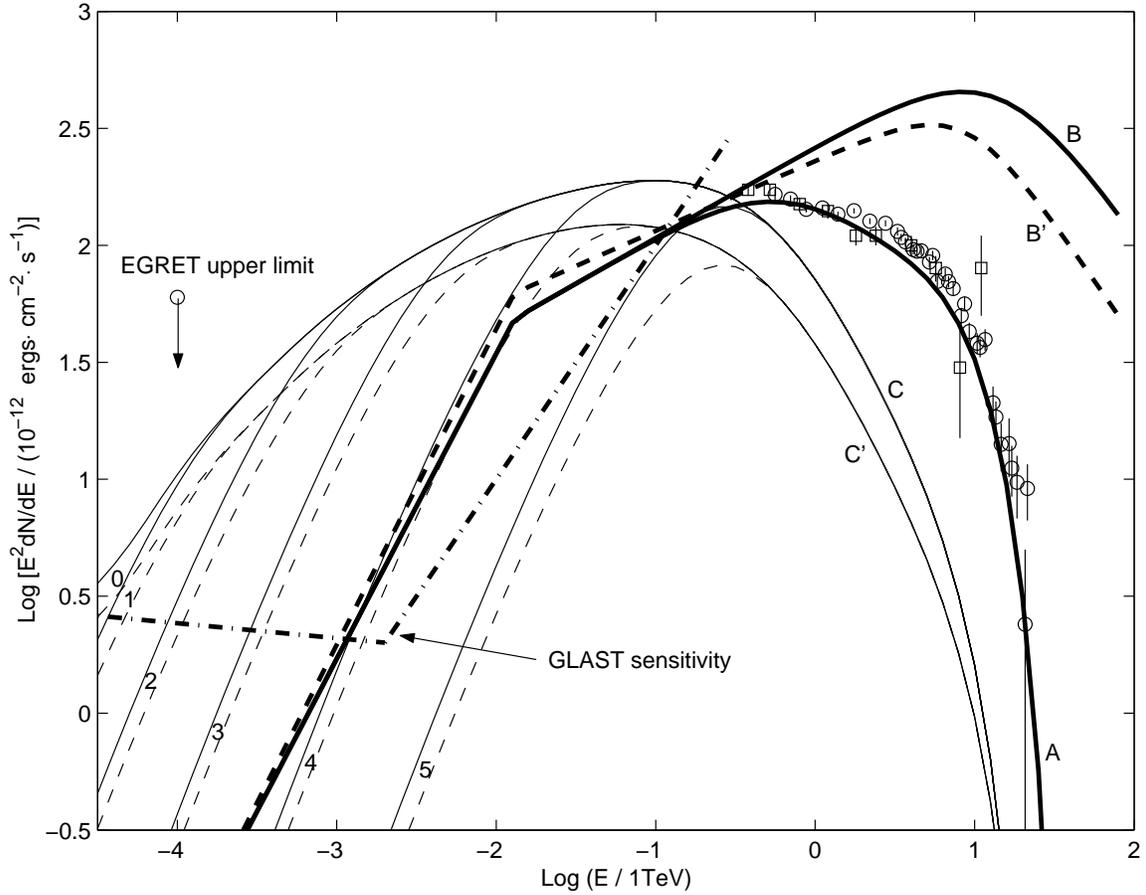,height=12.0cm}}
\caption{High-energy $\gamma$-ray spectra of Mrk 501 for a flare with 
$t_{\rm var}=0.5$ days and $\eta=1$. The spectral data are those measured 
by HEGRA ({\em circles}, Aharonian et al. 1999, 2001) and Whipple
({\em squares}, Krennrich et al. 1999), and their fitting is shown
by line A. The EGRET upper limit at $100$ MeV is from Catanese et
al. (1997). Thick solid (B) and dashed (B') lines are synchrotron
self-Compton intrinsic spectra, derived from the observed spectrum
A after correction for absorption on the IR background
``fast-evolution" and ``baseline" models of DS, respectively.  
Thin lines (C and C') are the secondary photon
spectra calculated in the text from the resulting pairs
interacting with the CMB. The thin lines labelled with numbers
$0-5$ correspond to intergalactic magnetic fields of zero,
$10^{-20}$, $10^{-19}$, $10^{-18}$, $10^{-17}$, and $10^{-16}$ G,
respectively. The thick dot-dash line represents the GLAST
sensitivity computed for an exposure time of 0.5 days.}
\label{fig1}
\end{figure}

\begin{figure}
\centerline{\psfig{file=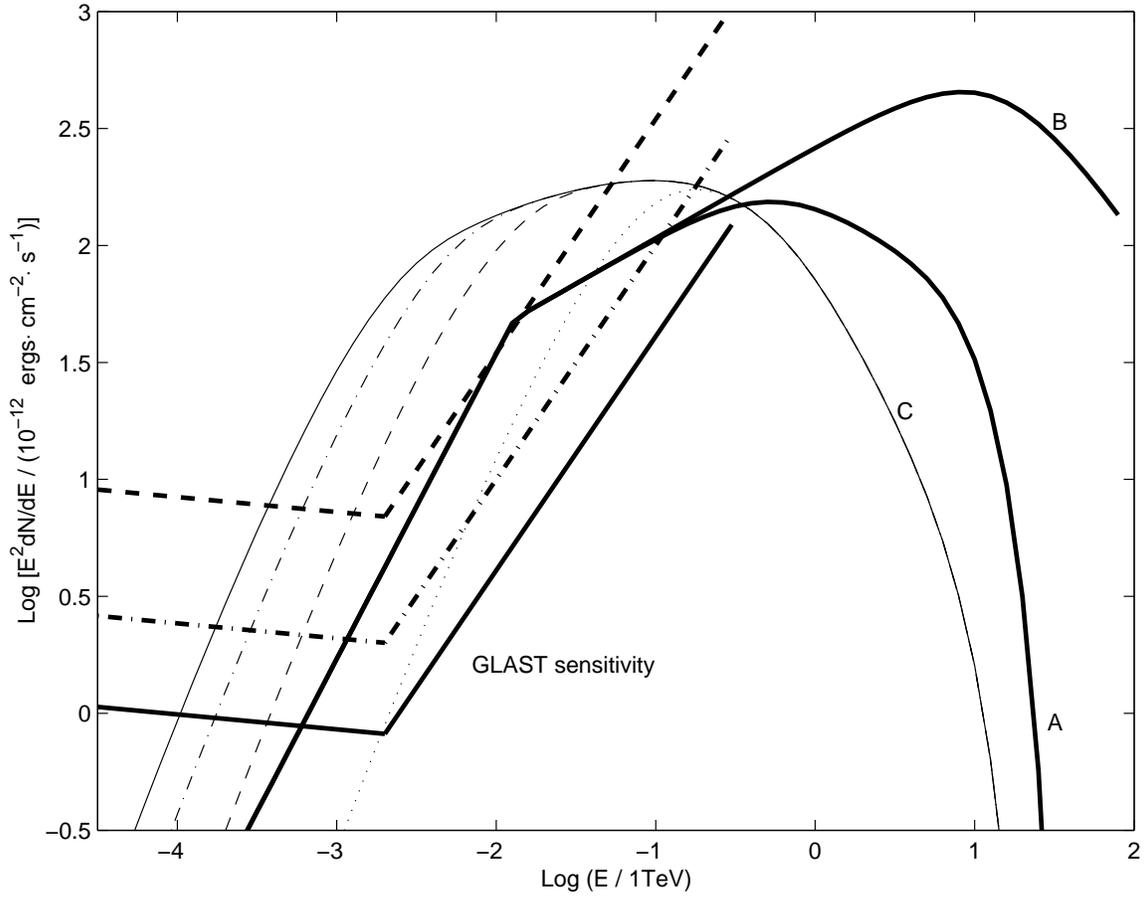,height=12.0cm}} \caption{Same as
Fig.1 but for different flare durations and ``fast-evolution" IR background.
The predicted signals are thin lines while three thick lines are
the GLAST sensitivity, both for a given integration time. The  solid,
dot-dash and dashed lines correspond to $t_{\rm var}=3$ days, 0.5
days and 1 hour for $B_{\rm IG}=10^{-18}$ G. The thin dotted
line is the signal for $t_{\rm var}=3$ days and $B_{\rm IG}=10^{-16}$ G.}
\label{fig2}
\end{figure}

\end{document}